# Nano antennas array for improving the light emission rate of organic phosphorescence

**Zahi Laty, Ofir Sorias and Meir Orenstain**

*Department of Electrical Engineering, Technion University, Haifa 32000, Israel*
*zahilaty@gmail.com

**Abstract:** In this paper we present how nano antennas arrays based on Al and Au were successfully increased the phosphorescence emission rate of Iridium-polypyrrole (Irppy3) and Platinum-octaethyl-porphyrin (PtOEP) solutions - which are commonly used as the active layer in organic light emitting diode (OLED) devices. Using electron beam lithography, the nano antennas were embedded inside those organic solutions and lab measurements were conducted. The lifetime of the organic solutions excitons', was decreased by 15%, while the photoluminescence was improved by 10% due to the improvement of the OLED internal and extraction efficiency.

## 1. Introduction

A common approach in the plasmonics research is the use of nano antennas which are basically metal nano particles (MNP) lying on or inside an insulator. Such particles have the ability of coupling between a far and a near field – just like any regular antenna does – but for optical wavelengths rather than microwave or radio wavelengths. In addition, those particles are also concentrating an electromagnetic field around them, in a similar way to what had been described in the Mie scattering theory, and therefore, they are also known as localized surface plasmon resonance (LSPR) [1,2].

By coupling between the far and the near field, and concentrating electromagnetic field to a small area, the particles can increase the efficiency of many devices, among them is the OLED [3,4] which often comprised of an emitting layer made of Irppy3 or PtOEP. Those combinations of metal and polymer are widespread because of their ability to harvest both singlet and triplet [5].

Although the OLED has many advantages, it possess poor efficiency, mainly due to total internal reflection (TIR) and a large non-radiative recombination in its phosphoresce emission mechanism which characterizes 75% of the excitons because of the 1:3 singlet-triplet ratio [5,6]. The reason for this large non-radiative recombination, is the fact that the exciton life time of the phosphoresce emission is about 1us, which is relatively long (compare with the 10-100ns life time of the fluorescence emission) and therefore, if the phosphoresce lifetime could be decreased - than the OLED internal efficiency might be improved.

In this paper we examined the phosphorescence from an optically pumped OLED made of either Irppy3 or PtOEP. Using a finite difference time domain (FDTD) simulations and lab experiments, we studied the effect of a nano-antenna array on the OLED device, and unlike previous studies that have been done in this field such as [3,4], in this study, the particles are arranged in a specific order - the antenna’s array, and are not scattered randomly in the active material as in self-assembled techniques such as thermal evaporation.

## 2. Simulations and design

### *2.1 Background and theory*

The basic model which is illustrated in figure 1 includes an organic material on an Indium Tin Oxide (ITO) layer, confined by two infinite glass sheets. The organic material is either Irppy3 or PtOEP, and it contains the nano antennas array. An incoming plane wave hits the structures and excites LSPR modes between the antennas' gap (i.e., the volume between the pair of metals, which is somehow analogous to the electric circuit in radio frequency antennas), and since we want to imitate the classic $\lambda/4$ dipole antenna concept, the plane wave polarization is in parallel to the antennas' long dimension.
In order to examine the ability of changing the antenna's resonance frequencies to match the PL spectrum of the OLED active layer (510nm for Irppy3 and 650nm for PtOEP), we used the local field enhancement inside an antenna gap as a figure of merit. This figure of merit is based on the idea that as the incoming plane wave propagates through the structure, some of it is absorbed (due to the complex permittivity of the organic layer and of the metals) and when the absorption is in the area around the antennas and in their gap, a local field is formed (this field is the result of the metal's electrons moved by the plane wave to the boundaries). When practicing the Fermi golden rule to our model [1], one can see that the areas in which the local field enhancement is formed, have a larger photon's density of states (DOS), and therefore, the internal efficiency is increased in these areas. The amount of field enhancement in the center of the antennas' gap was found as a function of the wavelength, and an optimization of the structure was done to maximize its enhancement values in corresponding to the PL spectrum.

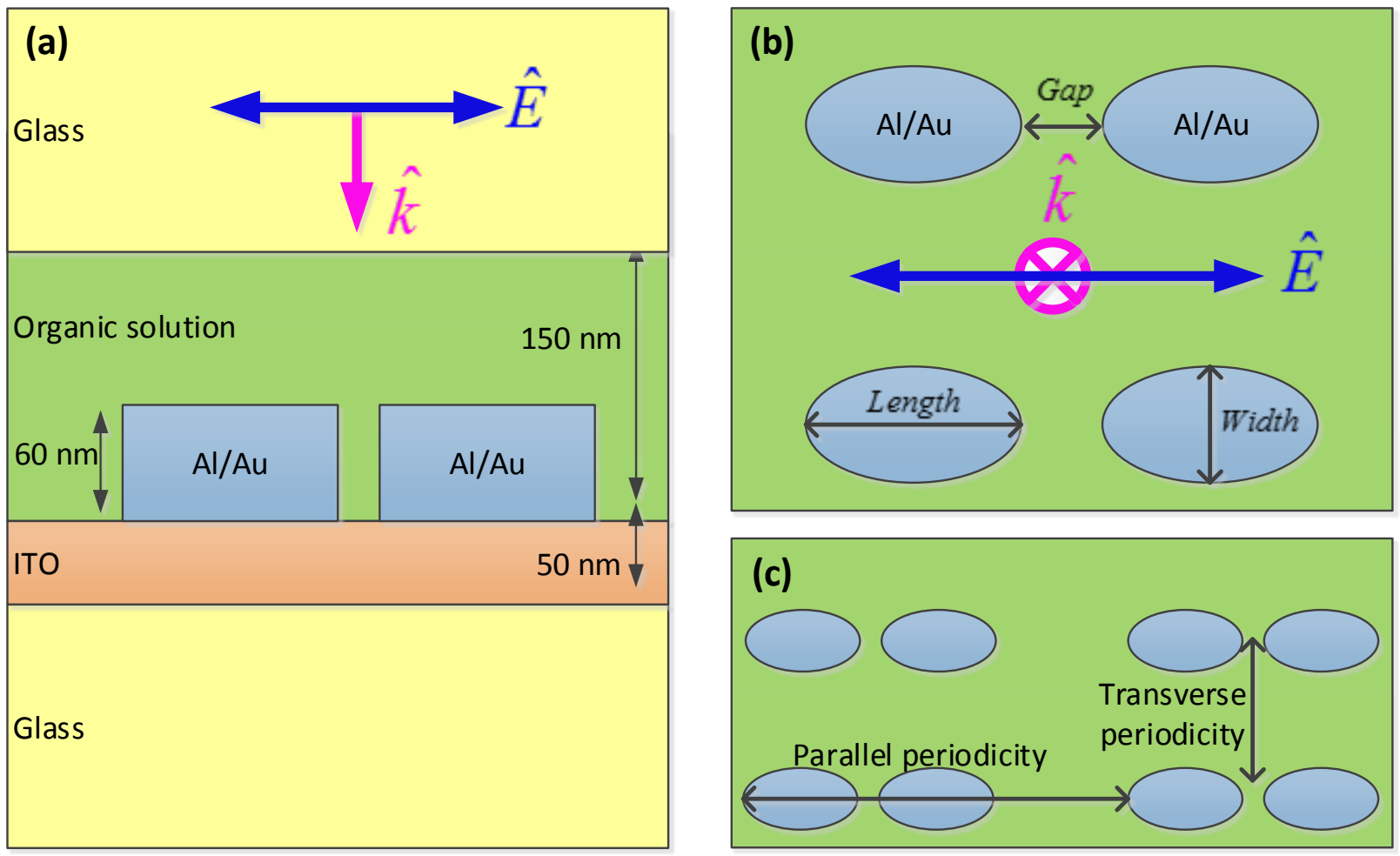

Fig. 1. The 3D simulated FDTD model: a periodic array of metal nano particles inside an organic film (green), illuminated by a normal propagating (purple arrow) plane wave with polarization illustrated by the blue arrows. (a) A yz plane slice of the structure. (b) A top view of a xy plane slice. (c) Peripheral top view of a xy plane slice.

Please note that from here and onward we will assume that the antenna shape is an ellipse with constant thickness equal to 60nm (as can be seen in figure 1).

### *2.2 The influence of the particle size, thickness, and gap*

As for the particle length, in order to have a resonance at a short wavelength, short antennas are required. However, when the antennas are shorter the antennas' cross-section (i.e. the ability of the antenna "to attract" a photon) decrease, so the enhancement factors are decreasing as well, as can be seen in figure 3(a). This trade-off is somehow reminding that classic radio-frequency antennas also have a dependence of the radiative efficiency on the antenna length - $\varepsilon_{rad} \propto l^2$ .

This concept is rising again when analyzing the effect of the antenna's width in figure 3(b). Again, shorter widths result in a smaller cross-section and enhancement factor (the resonance wavelength however, is not changing because it is determined only according to the dimension of the particle in the direction of the polarization). Of course that one need to takes into account the losses of the transverse polarization before choosing the optimal width.

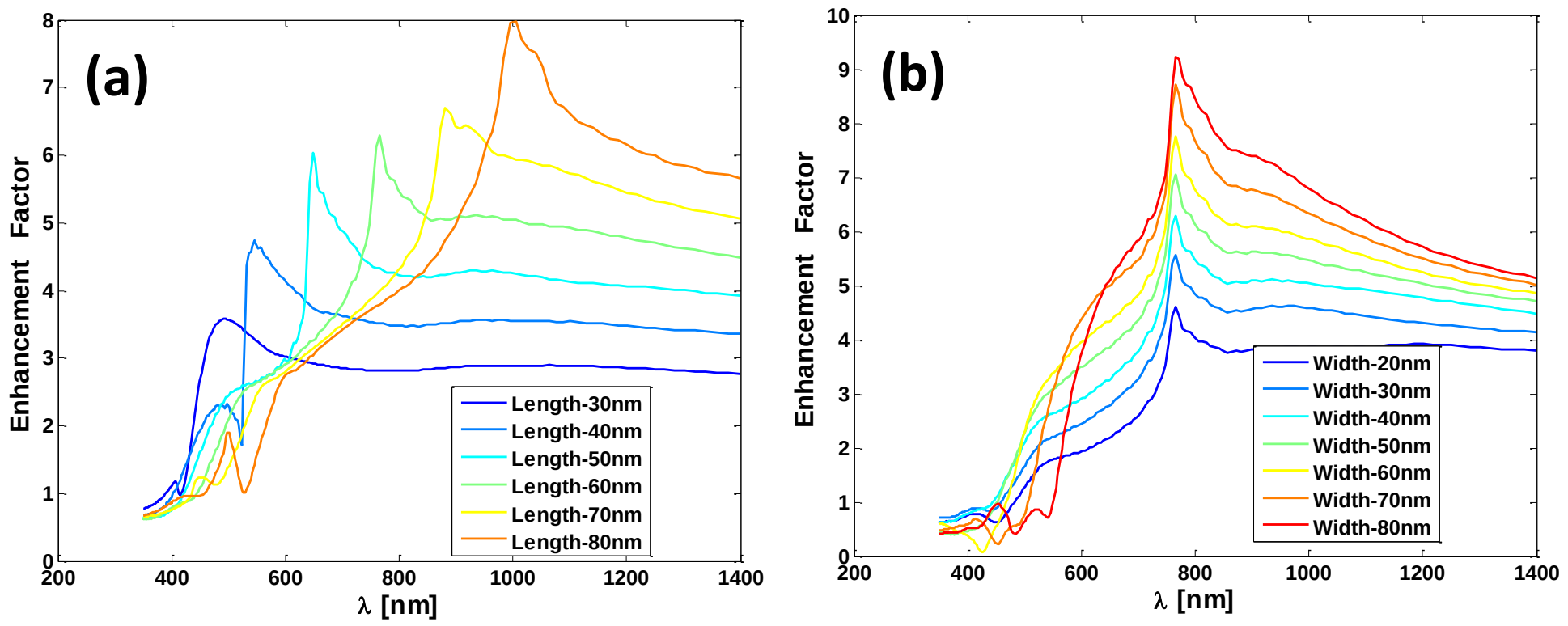

Fig. 2. The influence of the antennas' length (a) and width (b) on the maximum E field magnitude enhancement factor, obtained by calculating the field at the center of the ellipse Al antenna's gap inside Irppy3, in comparison to the field at the same spot of the reference model

As for the antenna's gap we found in our simulations that the smaller the gap is, the larger is the maximum field enhancement. Although it is true that by increasing the gap length, one can have a larger area of enhancement, it should be remembered that if our goal is to decrease the excitons' lifetime – rather than just increasing the overall PL, than what should matter most is the maximum field enhancement.

*2.3 The influence of array periodicity*

When looking at antennas with the same size but different array periodicity in the axis that is transverse to the plane wave polarization such as in fig 4(a), we can see the "array factor" effect; As we increased the distance between the antennas, each antenna is collecting more power than before, and the field enhancement is increasing as well. However, when increasing the distance, we have less antennas in a given area, so the overall effect is decreased – in the same proportion as the distance increment.

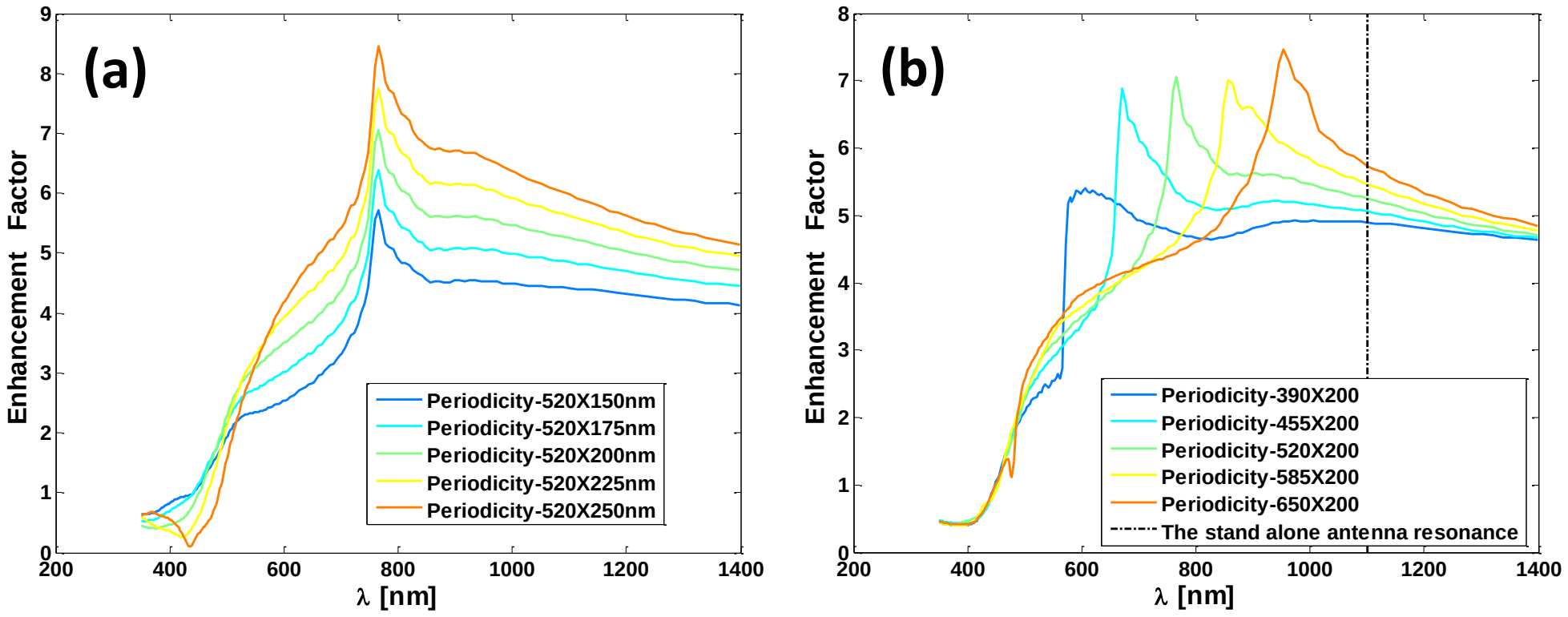

Fig. 3. The influence of the antennas' transverse (a) and parallel (b) periodicity on the maximum E field magnitude enhancement factor, obtained by calculating the field at the center of the ellipse antenna's gap inside Irppy3, in comparison to the field at the same spot of the reference model.

The analysis of the parallel periodicity, i.e., the periodicity on the main axis of a dipole antenna is more complicated because it is adding another resonance to the problem. As can be seen in fig 4(b), when increasing the periodicity, even without changing the antenna's size at all, we get the resonance shifted almost by the same ratio as the periodicity increase. For example, it is possible to calculate the ratio between the red and blue series in the figure: the periodicity ration is ( $650/390=1.66$ while the field enhancement ratio is $9.54/5.79=1.64$. However, this effect is limited at the shorter wavelengths (no matter how much the resonance is shifted, still we get no enhancement for the shorter wavelengths).

The primary reason for that is because the measured enhancement is a function of two parameters – the periodicity and the stand-alone antenna, and each one of them is contributing its own resonance and becomes dominant in a different spectrum.

Because of the fact that we have two different resonance mechanisms, changing the parallel periodicity parameter can assist us in creating the effect of a "smaller antenna" and therefore, to shift the resonance according to our needs. For example, in the simulations a 120nm ellipse with a 20nm gap was used. Considering the $\lambda/4$ rule and the fact that the Irppy3 refractive index is about 1.7, the stand-alone antenna should have had a resonance at 884nm because $1.7\cdot 2\cdot(120+120+20)=884nm$. However, when we used the antenna array periodicity impact, we managed to shift the resonance and to get reasonable enhancement factors in the relevant PL wavelength - $\lambda \approx 509nm$.

The ability to shift the resonance also depends on the antenna's length, because the periodicity resonance effect is more significant in shorter wavelengths and when we are far from the stand-alone antenna resonance.

### *2.4 The optimal array with constrains*

After considering all the parameters that were mentioned above, we found that the optimal ellipse antenna for optically-pumped OLED is made from Al and has the following dimensions: 120nm length, 80nm width and 60nm thickness. For an OLED with an Irppy3 film, the best unit cell periodicity is 650 nm for the axis of the long radius and 160 nm for the shorter one. When using PtOEP, the best periodicity is 390 on 160 nm. The field enhancement profiles of this antenna in PtOEP can be seen in fig 5.

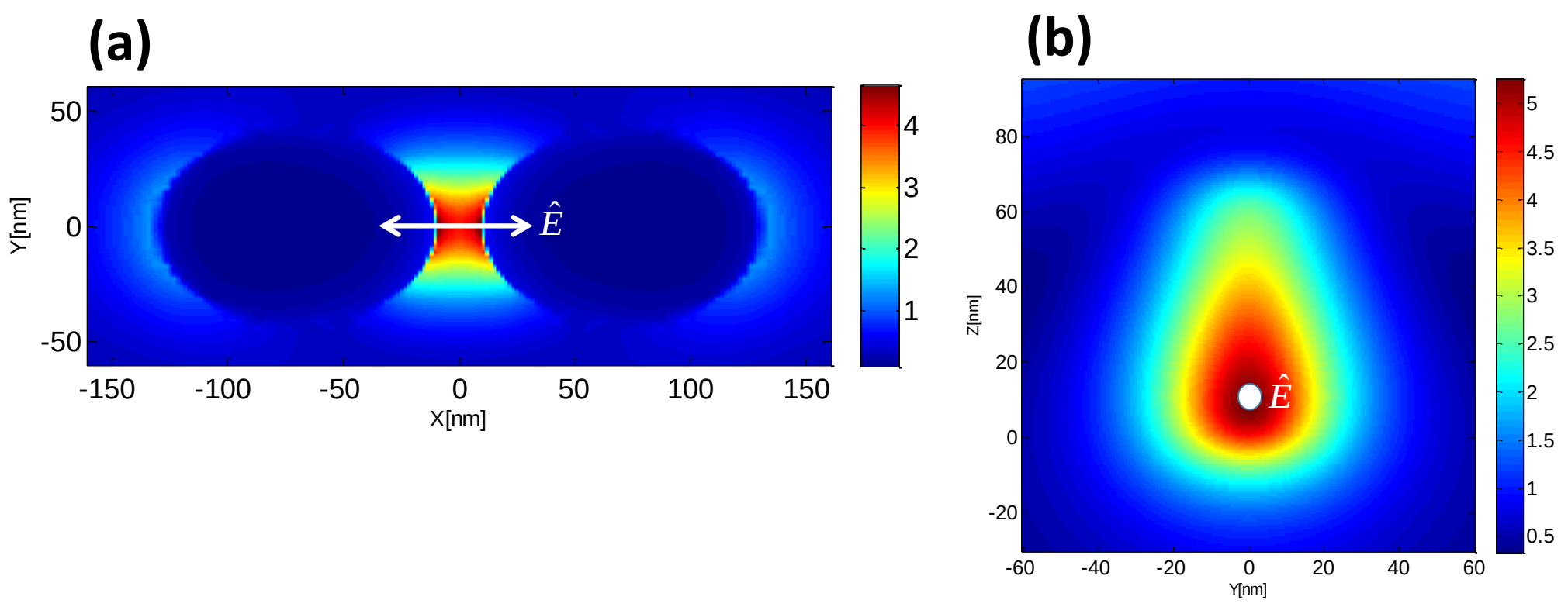

Fig. 4. A XY (a) and YZ (b) slice of an E Field magnitude enhancement, obtained by a simulation of the optimal ellipse antenna inside PtOEP at 652[nm]. The color bar represents the E field magnitude enhancement factor compared to that of the incident wave.

### 3. Fabrication and experimental details

The fabrication process of the devices that was described in section 2 is depicted in the following figure:

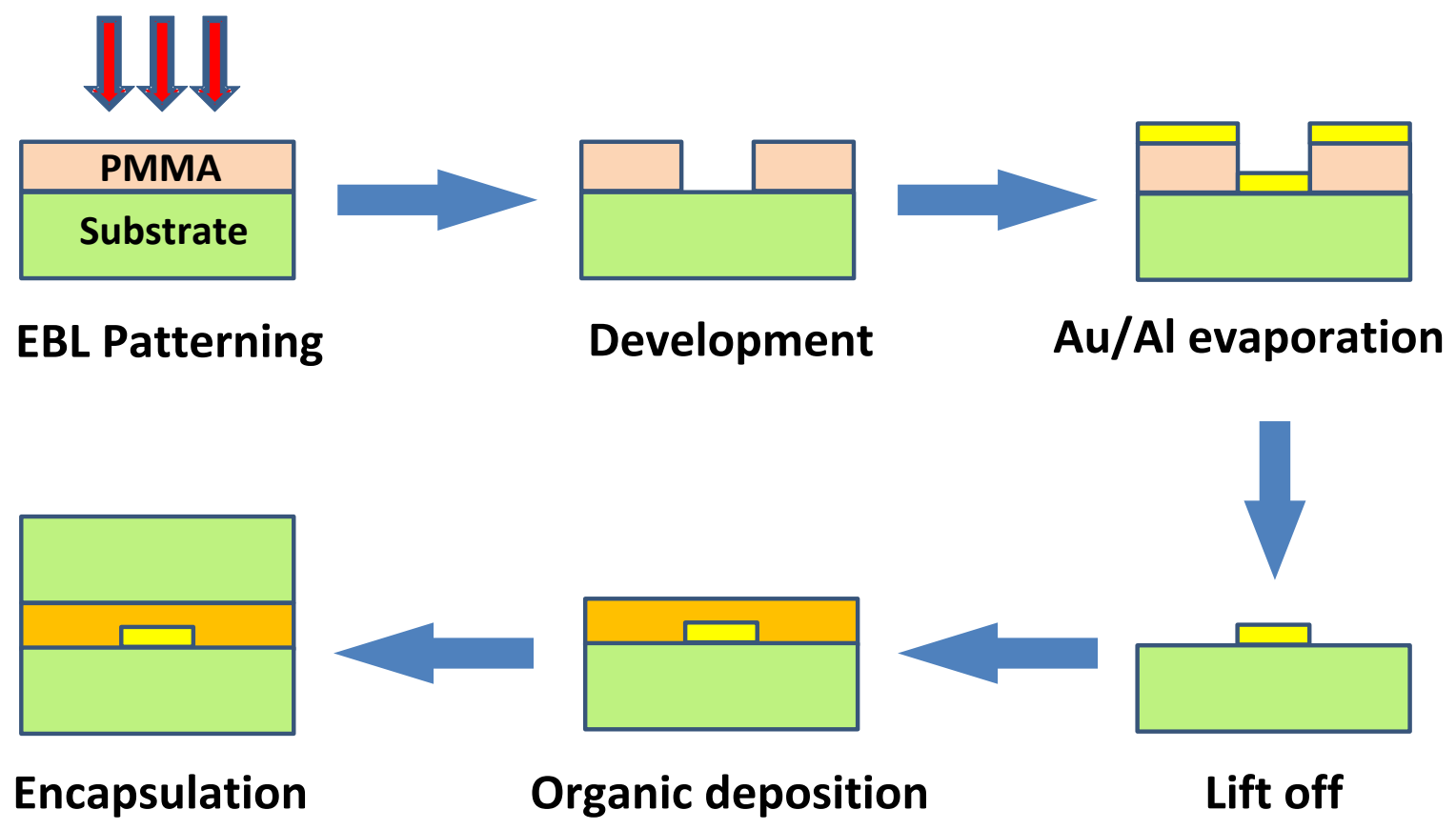

Fig. 5. Fabrication process guidelines.

As can be seen in figure 5, the first step in the fabrication process is preparing a glass substrate with a thin layer (of about 30-50 nm) of ITO and polymethyl methacrylate (PMMA) for EBL patterning. The PMMA is a spin-coated photo-resist that is sensitive to electrons, and it will undergo a chemical change when exposed to the EBL.
Then comes the EBL patterning, where a previously designed model is loaded, and the EBL machine is exposing the PMMA corresponding to the desired geometries and sizes.
The second step is the development in which the exposed parts of the PMMA are removed by a chemical etching process.
The third step is the evaporation of the desired metal (Au or Al around 60 *nm* thicknesses for this work) on the remaining PMMA or the exposed glass.
The fourth step is the lift-off process, in which the remaining PMMA, with the metal that was evaporated on it, are removed, ending up with the metal only where the MNPs were designed to be.

The fifth step is the organic deposition, in which an organic solution, around 150*nm* in thickness, is deposed on the substrate and between the MNP gaps, leading to the final step where the encapsulation of the closing glass is done, and the device is ready for the experiment.
Although the fabricated devices are quite similar to the designed ones, they are never identical and the dimensions might be different up to +/-8nm.

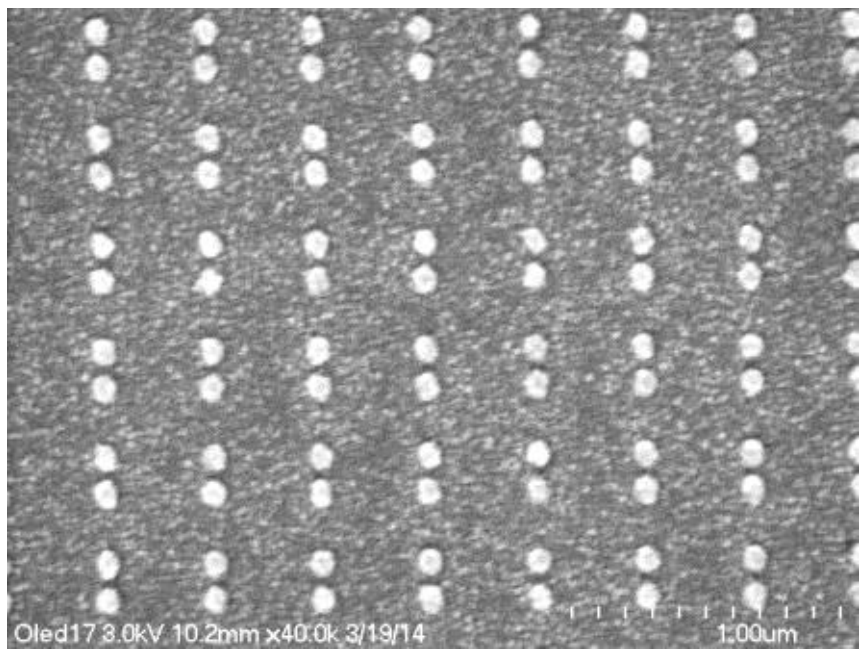

Fig. 6. A SEM photo of a device. The array periodicity is 390nm for both axes, while the size of each ellipse is approximately 100X80nm, and the length of the gap is 20nm.

## 4. Results and conclusions

### *4.1 Experimental details*

Antennas arrays with different dimensions were fabricated on a glass substrate. Each array was made of 122X122 ellipse pairs and with a periodicity of 400nm for both axes, so the total size of a device is about 50X50um.

A 375nm or 405nm (depend on the organic solution's) pulsed laser diode with a spot size of about 10X10um excited the device and initiated the fluorescence and phosphoresce emission. In order to image the emitted light, an X40 objective was used, mounted on a Ti-U Nikon microscope, with a 442nm dichroic mirror to avoid from collecting the reflections of the laser beam from the OLED. The emitted light was then coupled to an optic fiber and into an Edinburgh FN-980 system, where it was filtered by a monochromator and measured with visible detectors. This setup scheme is illustrated in fig 7.

In the PL experiment we measured the intensity of the light that was emitted from the device for each wavelength between 450nm to 750nm in compared to a reference device without nano antennas. The light emission enhancement factor was noted for the wavelength in which the solution emission is the largest – 509nm for Irppy3 and 650nm for PtOEP.

In the lifetime experiment we tuned the monochromator to these emission's peak wavelengths, and using a 10us pulse, that triggered a single photon detector, we built a histogram of the detections count per time unit that followed the pulse. From this histogram we extracted the phosphorene lifetime of the exactions that were formed in the solutions in response to the pulse (note that those histograms are composed from both a fast fluorescence and a slow phosphorescence lifetimes). The ratio of the phosphorescence lifetime of a device with nano antennas, to the phosphorescence lifetime of a reference device was noted.

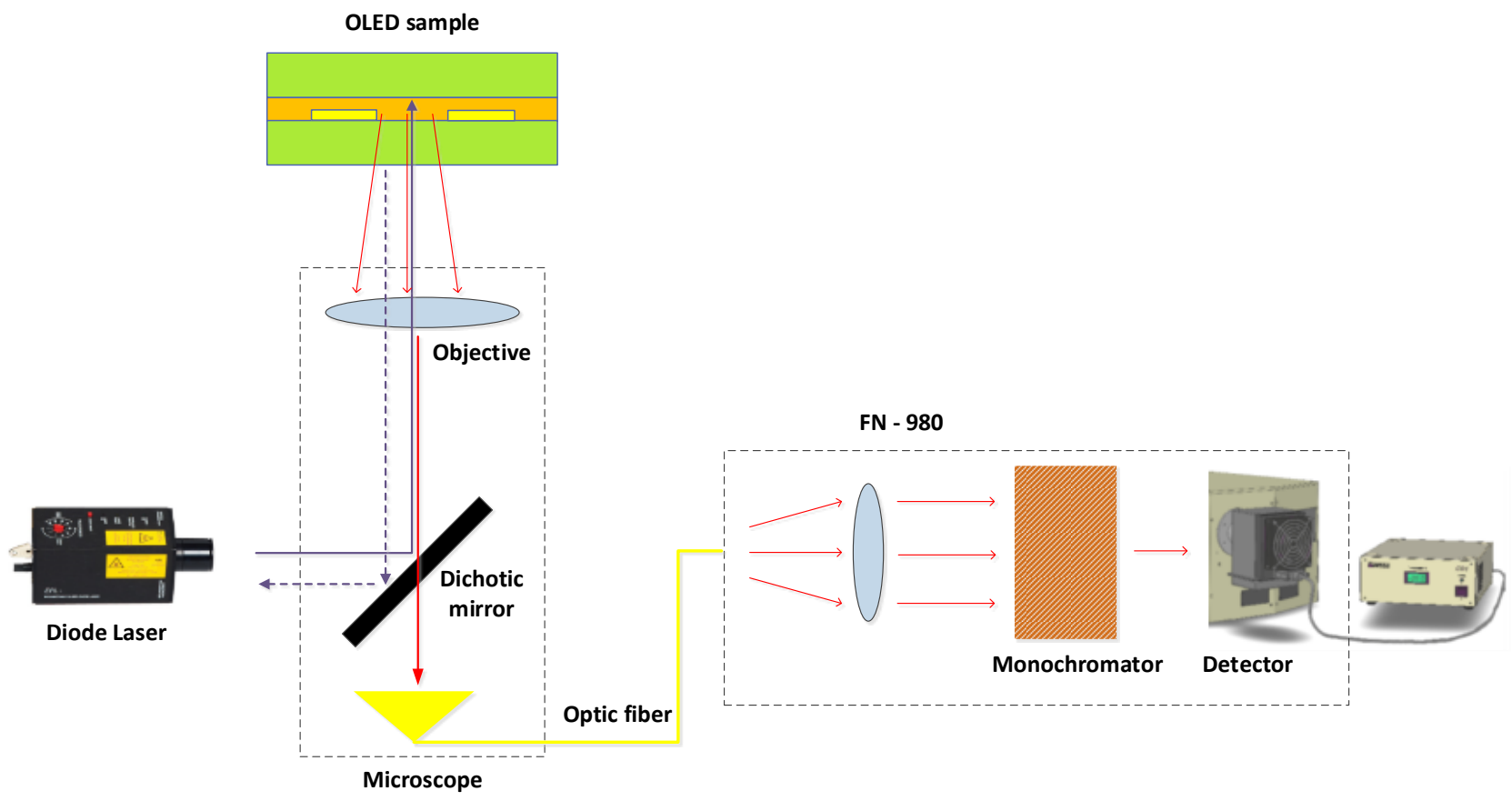

Fig. 7. The experiment setup scheme.

*4.2 Results and conclusions*

The measured PL enhancement factor and the extracted lifetime ratio of 40 devices that contained different antennas and organic solutions, are presented in the scattering figure below. Each dot represent a different device and its location in compare to the black square that represent a reference device, indicates the PL and lifetime improvements (or worsen). The most desirable device will be the one that has the largest lifetime decrease and the largest PL increase (i.e. it is located in the bottom-right area of figure 8), while the most undesirable device will be the one that has a PL decrease without any lifetime improvement at all (i.e. it is located in the top-left area of figure 8).  Of course, any dot that is located more to the right and downwards compared to the black reference square is still considered as a success (meeting our research goal).

The first conclusion that can easily be noticed is the existence of an inherent trade-off. The lifetime decrease is almost always coming at the expense of the PL performance, so actually, no device can be located in the far bottom-right or top-left areas. In the dashed line, we marked only the best configurations which are able to provide good lifetime decrease factor for a certain PL enhancement value.  Please note that we omitted two devices from this selection, because they seemed for us "too good to be true". This line can give any future researchers in the OLED plasmonics field a reference for what could be achieved with a proper design.

The second conclusion is the strong compliance between the simulations and experiments. One can notice that the devices that compose the dashed line have an identical or very similar dimensions (with a slight changes of +/-10nm) to those that we have found in section 2 to be optimal. This fact indicates a strong correlation between the hypothesis and the results.

However, one can claim that based on the field enhancement factors that were shown in the simulations, we could have been expected for lifetime decrease factors of about 75%, rather than 5%-15% (giving into account the fermi golden rule and the increase in the photonic DOS). The main reason for this discrepancy is that our antennas had influenced mainly on the emitting dipoles between the gaps, while the other surrounding area remained unchanged, so only a small part of the emitters that were included in the measurements were actually enhanced (the size of the enhanced volume is only about 0.5% - 0.4% out of the entire emitting volume).

The last conclusion that could be discussed here is the devices' variance. It can be noticed that the Irppy devices (the blue dots) are characterized by a small variance and quite a similar performance, while the devices with the PtOEP and Au (the red dots) or Al (the green dots) are more scattered. We believe that the reason for this is that the Irppy PL spectrum is wider than the PtOEP PL spectrum, so differences in the Irppy emission are less observable - but this is only an assumption without any proof so far.

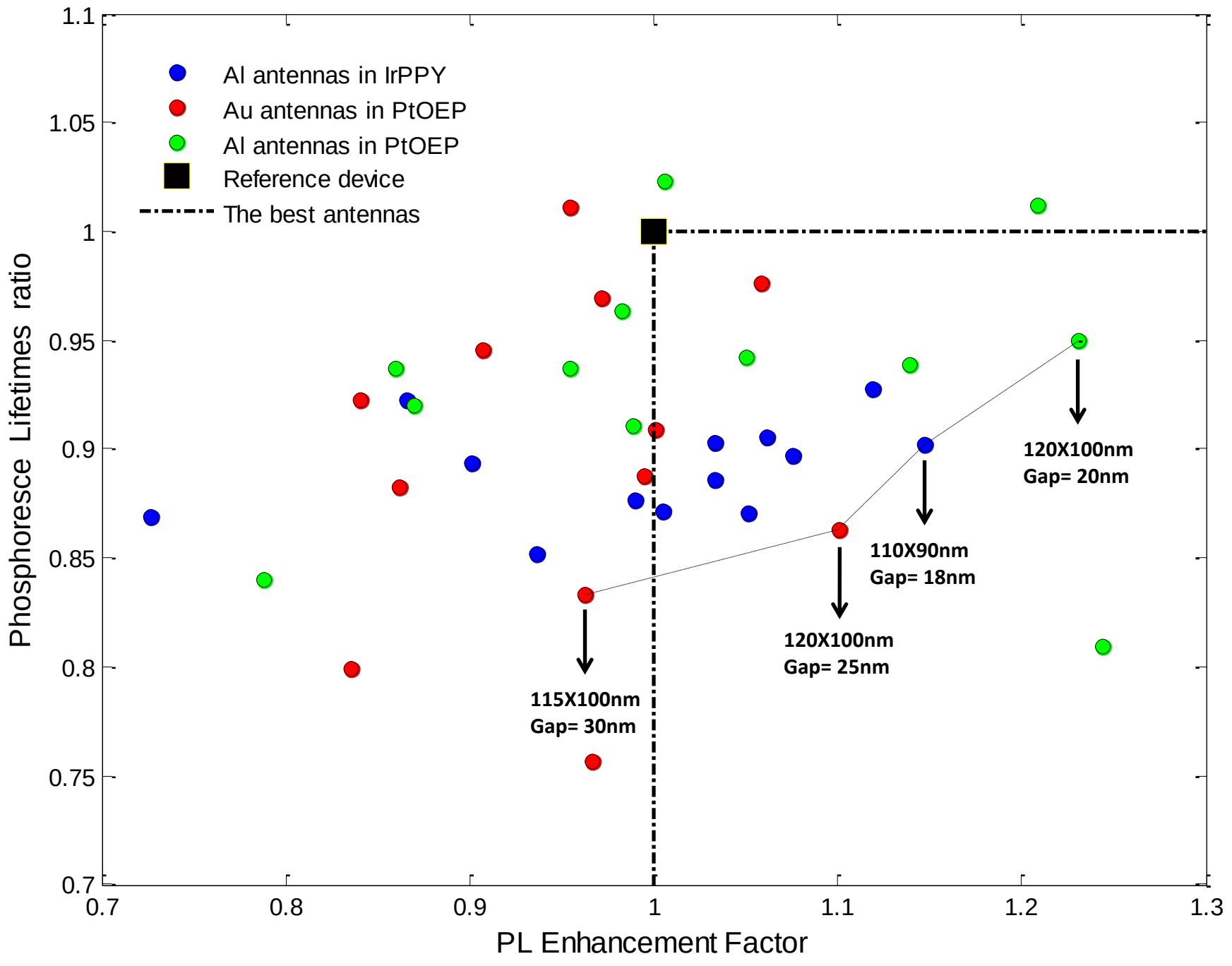

Fig. 8. A scatter plot of 41 devices, where x-axis represent the PL enhancement factor and y-axis represent the lifetime increase factor in compare to a reference devices without nano-antennas. 15 out of the 40 devices (placed at the bottom-right area of the figure) meet all of the research goals and objectives.

## 5. Summary

In this paper we found a set of optimal configurations for a nano-antennas array in Irppy3 and PtOEP solutions, than can improve the OLED internal and extraction efficiency and proved a strong compliance between our simulations methods and the experiments.

These optimal configurations have a reasonable dimensions (i.e. that can be fabricated with a standard EBL processes) thanks to some novel design concepts that we have found such as the antenna array periodicity impact.

## Acknowledgments

We would like to thanks Prof. Nir Tessler and Dr. Olga Solomeshch for their significant aid in all of the aspects that are related to the organic solution and the fabrication of the devices.

## References and links